# Identity Based Strong Designated Verifier Parallel Multi-Proxy Signature Scheme

**Sunder Lal and Vandani Verma**
**Department of Mathematics, Dr. B.R.A. (Agra) University**
**Agra-282002 (UP), India**
**E-mail- sunder_lal2@rediffmail.com, vandaniverma@yahoo.com**

**Abstract-** This paper presents a new identity based strong designated verifier parallel multi-proxy signature scheme. Multi-Proxy signatures allow the original signer to delegate his signing power to a group of proxy signers. In our scheme, the designated verifier can only validate proxy signatures created by a group of proxy signer.

## 1. INTRODUCTION

Shamir [8] in 1985 introduced the concept of identity(ID) based cryptosystems where, a users public key is derived from his identity and the corresponding private key is generated by a trusted third party called the *Private Key Generator* (PKG). ID based cryptosystems are advantageous over the traditional public key cryptosystems (PKC) as they avoid the need of certified public key register.

Jakobsson et al [4] presented the concept of designated verifier signatures (DVS) in 1996. In DVS, the signer specifies a designated verifier who can only determine the validity of the signatures. However, the verifier in general is not able to convince other parties on the validity of the signatures, because he himself is able to produce the indistinguishable signatures. Sadeenia et al [6] added the concept of strongness in DVS that forces the designated verifier to use his secret key at the time of verification. When only two (unknown to each other) users can verify the signatures, the scheme is said to be bi-designated scheme [7].

In day-to-day life, many legal documents require signatures from more than one party. To meet these requirements, cryptography provides a mechanism known as multi-signatures proposed by Itakura et al [3] in 1983. A multi-signature provides multiple signers to generate a valid signature for a single message. Based on the nature of applications, the multi-signatures are categorized into two types: serial multi-signature and parallel multi-signature. In serial multi-signature, a signer signs the message and sends it to the next signer for further processing; the next signer after verifying his predecessor's signature, signs the received components. The serial multi-signature generation is considered to be complete when the last signer signs the message. In case of parallel multi-signature, the signature of each signer is carried out on the message itself but not on the signatures of the other signers. In order to complete the parallel multi-signature generation, a designated clerk combines all the individual signatures after verifying them.

Proxy signatures proposed by Mambo et al [5] is a variation of normal signature schemes, in which an original signer delegates his signing power to another signer called the proxy signer. The signature generated by the proxy signer is called the proxy signature for the original signer. But in some practical applications, the original signer may delegate his signing power in a distributive manner to all members of a group of specified proxy signers ensure individual accountability of each participant signer. The proxy signatures are obtained by combining (serially or parallel) all such signatures. Such a signature is called multi-proxy signature scheme. This was first proposed by Hwang et al [2] in 2001. The signature generated by the specified proxy signer is called the multi-proxy signature for the original signer.

In this paper, we propose an ID based strong designated verifier parallel multi-proxy signature scheme. Our scheme is the based on the ID based multi-proxy signature scheme proposed by Chen et al [1]. In the proposed scheme, the designated verifier can only verify the multi-proxy signature generated by a group of proxy signers and he cannot convince any third party about the validity of the signatures. To the best of our knowledge there is no existing scheme on this concept.

The rest of the paper is organized as follows: some definitions and preliminary works are given in section 2. Section 3 contains the model of the proposed ID based strong designated verifier parallel multi-proxy signature (ID-SDVPMPS) scheme and in section 4 we propose the ID-SDVPMPS. The security and the computational efficiency of the scheme is discussed in section 5 and 6 respectively. Finally, section 7 concludes the paper with applications.

## 2. SOME DEFINITIONS

In this section, we define bilinear pairings, various Diffie-Hellman problems and gap Diffie-Hellman group.

### 1) Bilinear pairings

Let $G_1$ be a cyclic additive group with generator $P$, whose order is a large prime number $q$ and $G_2$ be a cyclic multiplicative group with the same order q. Let $e: G_1 \times G_1 \rightarrow G_2$ be a map with the following properties:

*Bilinearity*: $e(aP, bQ) = e(P, Q)^{ab}$ $\forall P, Q \in G_1$ and $a, b \in Z_q^*$.

*Non-degeneracy*: $\exists P, Q \in G_1$, such that $e(P, Q) \neq 1$, the identity of $G_2$.

*Computability*: There is an efficient algorithm to compute $e(P, Q)$ $\forall P, Q \in G_1$.

Such pairings may be obtained by suitable modification in the Weil-pairing or the Tate-pairing on an elliptic curve defined over a finite field.

### 2) Gap Diffie-Hellman Group

*Discrete Logarithm Problem (DLP)*: Given $Q \in G_1$, find an integer $a \in Z_q^*$, such that $Q = aP$, $P$ is a generator of $G_1$.

*Decisional Diffie-Hellman Problem (DDHP)*: Given $P, aP, bP, cP$ in $G_1$, decide whether $c = ab \bmod q$.

*Computational Diffie-Hellman Problem (CDHP)*: Given $P, aP, bP$, compute $abP$

*Bilinear Diffie-Hellman Problem (BDHP):* Given $P, aP, bP, cP$ compute $e(P, P)^{abc}$.

*Gap Diffie-Hellman group (GDHP)*: A class of groups, where DDHP can be solved in polynomial time but no probabilistic algorithm exists that can solve CDHP in polynomial time.

## 3. MODEL FOR THE PROPOSED ID-SDVPMPS SCHEME

In this section we define the phases through which our scheme is generated. Our scheme has five phases described as follows:

- **Setup:** Given security parameter $k$, this algorithm outputs the public parameters.
- **Key generation:** Given a user identity and the public parameters, this algorithm computes secret key of the user and the public key of the user's.
- **Proxy key generation:** Given original signer's secret key, secret key of the original signer group, public key of the designated verifier, warrant on message '$m$' and some random numbers, it outputs the proxy key of each proxy signer.
- **Multi-Proxy Signature Generation:** Given proxy signing keys of the signing group, random numbers, public key of the designated verifier and warrant on message '$m$', it outputs the multi-proxy signature on message '$m$'.
- **Multi-Proxy Signature Verification:** Given designated verifier secret key, signature $\sigma$ on message '$m$' it outputs whether $\sigma$ is rejected or accepted.

## 4. PROPOSED ID-BASED STRONG DESIGNATED VERIFIER PARALLEL MULTI-PROXY SIGNATURE SCHEME

The proposed scheme involves five roles: the private key generator (PKG), the original signer Alice, a set $PS = \{P_1, P_2, \ldots, P_n\}$ of proxy signers, a clerk Bob $\in \{P_1, P_2, \ldots, P_n\}$, and a designated verifier Cindy. It consists of the following five phases:

- **Setup**

  In this phase, PKG chooses a generator $P \in G_1$, a random number $s \in Z_q^*$ and computes $P_{pub} = sP$. PKG also chooses two cryptographic hash functions $H_1: \{0,1\}^* \to G_1$, and $H_2 : \{0,1\}^* \times G_1 \to Z_q^*$. The system parameters $(G_{1,}, G_2, P, P_{pub}, H_1, H_2, e)$ are made public and '$s$' is kept secret with PKG.

- **Key generation**

  A user '$U$' submits its identity $ID_U$ to PKG, which generates $S_{IDU} = sQ_{IDU}$ as the secret key and $Q_{IDU} = H_1 (ID_{IDU})$ is the public key of the user

- **Proxy key generation**

  To delegate the signing capability to proxy signers, Alice generates the signed warrant $m_w$ on message '$m$'. Each proxy signer generates the proxy key through the following protocol:

  - Alice chooses a random $r \in Z_q^*$ and computes $U = rQ_{IDC}$, $h = H_2(m_w // U)$, and $V = hS_{IDA} + U$. Alice sends $\sigma = (m_w, U, V)$ to each member $P_i$ of the signer group PS.
  - Each $P_i \in PS$ accepts the signatures $\sigma$ on $m_w$ iff $e(V, P) = e(Q_{IDA}, P_{pub})^h e(U, P)$. If valid, each $P_i$ computes the proxy key $S_{Pi}$ *as* $S_{Pi} = hS_{IDPi} + V$

- **Multi-Proxy Signature Generation**
  Each proxy signer $P_i$ generates the partial signature and an appointed clerk Bob, who is one of the proxy signers, combines the partial proxy signature to generate the final designated verifier multi-proxy signatures.
  - Each $P_i$ chooses $t_{Pi} \in Z_q^*$ and computes $Z_{Pi} = t_{Pi}Q_{IDC}$, $Z_P = \sum_{i=1}^{n} Z_{Pi}$, and broadcast $Z_{pi}$ and then each $P_i$ computes $H = H_2(m_w // Z_P)$, $X_{Pi} = HS_{Pi} + Z_{Pi}$. $(Z_{Pi}, X_{Pi})$ is the partial signature of $P_i$ on the message '*m*'. Each $P_i$ sends $Z_{Pi}$ to the clerk Bob.
  - Bob computes $Z_P = \sum_{i=1}^{n} Z_{Pi}$, $H = H_2(m_w // Z_P)$ and for each '*i*' verifies the correctness of partial proxy signatures $(Z_{pi}, X_{Pi})$ as $e(X_{Pi}, P) = e(Q_{IDPi} + Q_{IDA}, P_{pub})^{hH}\, e(Z_{Pi} + HU, P)$

    Once all the partial proxy signatures are found correct, Bob computes $X = \sum_{i=1}^{n} X_{pi}$. The valid multi-proxy signature on message '*m*' is $\sigma' = (m_w, Z_P, X, U)$.

- **Multi-Proxy Signature Verification**
  Cindy on receiving $\sigma'$ confirms the warrant $m_w$, computes $H = H_2(m_w // Z_P)$, and accepts σ as the valid multi-proxy signature iff

$$e(X - nHU, Q_{IDC})\, e(\sum_{i=1}^{n}(Q_{IDPi} + Q_{IDA}), S_{IDC})^{-hH} = e(Z_P, Q_{IDC})$$

- **Correctness**
  The verification of the multi-proxy signature is justified by the following equation:

$$e(X - nHU, Q_{IDC})\, e(\sum_{i=1}^{n}(Q_{IDPi} + Q_{IDA}), S_{IDC})^{-hH}$$

$$= e(\sum_{i=1}^{n}(X_{pi}) - nHU, Q_{IDC})\, e(\sum_{i=1}^{n}(S_{IDPi} + S_{IDA}), Q_{IDC})^{-hH}$$

$$= e(\sum_{i=1}^{n}(HS_{pi} + Z_{Pi}) - nHU, Q_{IDC})\, e(\sum_{i=1}^{n}(S_{IDPi} + S_{IDA}), Q_{IDC})^{-hH}$$

$$= e(\sum_{i=1}^{n}(H(V + hS_{IDPi})) + Z_p - nHU, Q_{IDC})\, e(\sum_{i=1}^{n}(S_{IDPi} + S_{IDA}), Q_{IDC})^{-hH}$$

$$= e(\sum_{i=1}^{n}(H(U + hS_{IDA} + hS_{IDPi})) + Z_p - nHU, Q_{IDC})\, e(-\sum_{i=1}^{n} hH(S_{IDPi} + S_{IDA}), Q_{IDC})$$

$$= e(Z_p, Q_{IDC})$$

## 5. SECURITY ANALYSES

- **Prevention of misuse**

  No proxy signer can use the proxy key for the purpose other than generating a valid proxy signature, because of the use of warrant $m_w$ in the signatures. He can only sign messages that have been authorized by the original signer.

- **Proxy protected**

  The original signer cannot create a valid multi proxy signature since each proxy key includes the private key $S_{pi}$ of each proxy signer.

- **Strong designated**

  The designated verifier Cindy uses his secret to check the validity of the signatures. Moreover, Cindy is not able to convince anyone else of the signatures. Hence, scheme provides the strongness property.

- **Strong Unforgeability**

  In our scheme, the clerk is one of the proxy signers but he has more power than other proxy signers. Assume that the clerk wants the proxy group to sign a fake message $m'$. He can change his $Z_{Pi}$ and therefore $Z_p$ can be changed but from the security of public one-way hash function $H_2$ it is impossible for the clerk to get $H'$ and $X'$ such that $(m', m_W, H', X', U)$ is a valid multi-proxy signature. Moreover, no user can forge the multi-proxy signature because he cannot obtain more information than the clerk.

## 6. COMPUTATION ASPECTS

In this section we compare the computational efficiency of the Chen et al [1] scheme and the proposed scheme. We will check that how many computations are required to add the property of strong designated verifier to the Chen et al [1] scheme.

**(Table 1)**

| Schemes | Proxy Key Generation | | | | | Multi-Proxy Signature Generation | | | | | Multi-Proxy Signature Verification | | | | |
|---|---|---|---|---|---|---|---|---|---|---|---|---|---|---|---|
| | H | M | E | P | I | H | M | E | P | I | H | M | E | P | I |
| 3 | 2 | 3 | 1 | 3 | - | 1 | 4 | 1 | 3 | - | 2 | - | 1 | 3 | - |
| Chen's | 2 | 3 | 2 | 3 | - | 3 | 2 | 3 | 3 | - | 1 | - | 3 | 2 | - |

**H** = Hash, **M** = Multiplication, **E** = Exponential, **P** = Pairing, **I** = Inverse.

**(Table 2)**

| Schemes | Comparison of Chen et al [4] and Scheme 3 |
|---|---|
| 3 | $5H+8M+3E+9P$ |
| Chen's | $6H+5M+8E+8P$ |

From the above two tables we can get the following conclusions:

- Only with the addition of one pairing we can add the concept of strong designated verifier to Chen et al [1] scheme.
- Moreover, Chen et al [1] scheme requires one hash and five exponential more than our scheme.

## 7. APPLICATIONS AND CONCLUSION

Country 'X' is set to start a new project for developing a nuclear weapon with the assistance of a group of 'n' scientists. During their operation, they encountered a problem that can only be solved by Cindy, who is a scientist from country 'Y'. Here 'Y' is the country which has already produced the same nuclear weapon and Cindy was one of the members of scientist panel. But, 'X' fears that if they discuss their concerns with Cindy, she may leak the news that 'X' is producing a nuclear weapon. In such situations, **Strong Designated Verifier Parallel Multi-Proxy Signature Scheme,** as proposed in this Chapter, can be used to generate a digitally signed document that is signed by all the scientists.

We proposed a new identity based strong designated verifier parallel multi-proxy signature scheme which is more efficient than ID-based multi-proxy signature scheme by Chen et al [1]. Our scheme has practical application in situations where the proxy-signature generated by specified group can only be verified by a single designated verifier.

## REFRENCES